\documentclass[runningheads]{llncs}
\usepackage[T1]{fontenc}
\usepackage{graphicx}
\usepackage{epsfig}
\usepackage{dsfont}
\usepackage{amsmath}
\usepackage{amssymb}
\usepackage{color}
\usepackage{boldline,multirow}
\usepackage[normalem]{ulem}
\useunder{\uline}{\ul}{}
\usepackage{siunitx}
\usepackage{booktabs}

\usepackage{cite}
\usepackage{cases}
\usepackage{hyperref}
\usepackage{bbding}
\usepackage{bm}
\begin{document}
\title{DeSAM: Decoupled Segment Anything Model for Generalizable Medical Image Segmentation}

\titlerunning{Decoupled Segment Anything Model}

\author{Yifan Gao\inst{1,2} \and Wei Xia\inst{2} \and Dingdu Hu\inst{1,2} \and Wenkui Wang\inst{3} \and Xin Gao\inst{2}\Envelope}

\authorrunning{Y. Gao et al.}

\institute{School of Biomedical Engineering (Suzhou), Division of Life Science and Medicine, University of Science and Technology of China, Hefei, China\\ \and
	Suzhou Institute of Biomedical Engineering and Technology, Chinese Academy of Sciences, Suzhou, China \and State Key Laboratory of Ultra-precision Machining Technology,Department of Industrial and Systems Engineering, The Hong Kong Polytechnic University, Kowloon, Hong Kong SAR, PR China \\ \email{yifangao@mail.ustc.edu.cn, hudingdu@mail.ustc.edu.cn} \\
	\email{Wang.wenkui@polyu.edu.hk} \\
	\email{xiaw@sibet.ac.cn, xingaosam@163.com}
}
\maketitle
\begin{abstract}
Deep learning-based medical image segmentation models often suffer from domain shift, where the models trained on a source domain do not generalize well to other unseen domains. As a prompt-driven foundation model with powerful generalization capabilities, the Segment Anything Model (SAM) shows potential for improving the cross-domain robustness of medical image segmentation. However, SAM performs significantly worse in automatic segmentation scenarios than when manually prompted, hindering its direct application to domain generalization. Upon further investigation, we discovered that the degradation in performance was related to the coupling effect of inevitable poor prompts and mask generation. To address the coupling effect, we propose the Decoupled SAM (DeSAM). DeSAM modifies SAM's mask decoder by introducing two new modules: a prompt-relevant IoU module (PRIM) and a prompt-decoupled mask module (PDMM). PRIM predicts the IoU score and generates mask embeddings, while PDMM extracts multi-scale features from the intermediate layers of the image encoder and fuses them with the mask embeddings from PRIM to generate the final segmentation mask. This decoupled design allows DeSAM to leverage the pre-trained weights while minimizing the performance degradation caused by poor prompts. We conducted experiments on publicly available cross-site prostate and cross-modality abdominal image segmentation datasets. The results show that our DeSAM leads to a substantial performance improvement over previous state-of-the-art domain generalization methods. The code is
publicly available at \url{https://github.com/yifangao112/DeSAM}.

\end{abstract}
\section{Introduction}
Deep learning models achieve remarkable performance in medical image segmentation when trained and evaluated on data from the same domain \cite{gao2023anatomy}. However, the generalizability of deep models may be poor to unseen out-of-domain data, which prevents the use of models in clinical settings. To mitigate the performance degradation caused by domain shifting, previous attempts focus on unsupervised domain adaptation \cite{ganin2015unsupervised} and multi-source domain generalization \cite{muandet2013domain}. However, unsupervised domain adaptation and multi-source domain generalization rely on training data from the target domain or from multiple source domains. Such requirements may not hold due to cost and privacy issues, making it difficult for real-world clinical applications. 

A more practical but challenging method is single-source domain generalization: using training data from only one source domain to train deep learning models robust to unseen data. The main solutions include input space-based and feature-based data augmentation \cite{guan2021domain} Yet there are some limitations with solutions: input space-based augmentation requires expertise to design the augmentation function, and feature-based augmentation usually requires complex adversarial training \cite{chen2022maxstyle}.

Compared to the above approaches, directly migrating models based on large datasets to medical image segmentation to improve generalization is an attractive approach. Some early work utilized pre-trained models on natural or medical images and achieved good performance \cite{chen2019med3d,zhou2021models}. However, due to the small capacity of the pre-trained models, the cross-domain generalizability of the deep models was not effectively improved.

Recently, vision foundation models have made great progress in image segmentation \cite{kirillov2023segment,zou2023segment}. Segment Anything Model (SAM) \cite{kirillov2023segment}, trained on more than 1 billion masks, has achieved unprecedented generalization capabilities on a variety of natural images. Some work shows that adapting SAM to medical image segmentation also shows satisfactory results \cite{ma2024segment,wu2023medical,zhang2023customized,li2023polyp}. These advances demonstrate the promise of training a powerful segmentation model with generalizability using pre-trained foundation models.

However, the prompt-driven SAM struggles with automatic segmentation without specific prompts, which hinders its application to domain generalization. Workarounds like using grid points or full-image bounding boxes as prompts can enable automatic segmentation but with significantly reduced performance compared to providing explicit prompts. We argue that the poor performance of fully automated SAM in medical image segmentation can be attributed to a mechanism, namely the coupling effect: image embeddings and prompt tokens interacted in the cross-attention transformer layer of the SAM mask decoder, which makes the final output mask highly dependent on the prompt. Therefore, even after finetuning, the model still tends to be more sensitive to wrong prompts (i.e., points not in the mask or the boxes significantly larger than the mask).

To address this issue, we propose Decoupled Segment Anything Model (DeSAM) in this work, a novel architecture for fully automated medical image segmentation based on SAM. We decouple the mask decoder of SAM into two subtasks: 1) prompt-relevant IoU regression, 2) prompt-decoupled mask learning. Specifically, we design two new modules and add to the fully automated SAM. The first one is the prompt-relevant IoU module (PRIM), which predicts IoU scores based on given prompt and generates mask embeddings. The second one is the prompt-decoupled mask module (PDMM), which fuses the image embeddings from the image encoder with the mask embeddings from PRIM to generate the mask. DeSAM minimizes the performance degradation caused by wrong prompts in the automatic mode.  Extensive experiments on two public datasets show that the DeSAM improves the robustness of fully automated prostate segmentation against distribution variations across different sites.

\section{Related Work}
\subsubsection{Single-source domain generalization}
Given training data from only one domain and generalizing to an unseen domain, single-source domain generalization is more challenging since there is less diversity in training domains. Chen et al. \cite{chen2020advbias} augment data with the random bias field, the common image artifact in clinical MR images. RandConv \cite{xu2021randconv} uses random convolutions for data augmentation. MixStyle adopts a combination of the style information of randomly-selected instances of different domains. Maxstyle \cite{chen2022maxstyle} expands the domain space with additional noise and the worst-case composition. Ouyang et al. \cite{ouyang2022csdg} propose a simple causality-inspired data augmentation method that greatly improves the cross-domain robustness of deep models. Unlike previous methods, DeSAM enhances the generalization ability of deep models without a complex pipeline for data augmentation, making it more competitive in practical applications.

\subsubsection{Segment Anything Model}
The remarkable extension ability of Transformer makes it possible to construct large-scale models with billions of parameters. SAM was the first proposed foundation model for image segmentation and has been applied to various computer vision applications \cite{cen2024segment,ren2024segment}. Intrigued by its unprecedented performance in natural image segmentation, considerable efforts have been made on its extended applications in medical image \cite{roy2023sam,gao2024mbanet}. In particular, some work attempts to adapt SAM to medical image segmentation, including fine-tuning mask decoder and image encoder \cite{ma2024segment,zhang2023customized,li2023polyp}. However, these approaches mainly focus on adaptating SAM for specific medical tasks, and may not fully address the challenges in automatic segmentation scenarios. Our proposed DeSAM tackles the performance degradation in automatic segmentation through a novel decoupled architecture. Unlike previous methods, DeSAM introduces prompt-relevant IoU module (PRIM) and prompt-decoupled mask module (PDMM) to leverage generalizable knowledge while minimizing sensitivity to poor prompts. Furthermore, DeSAM's multi-scale feature fusion enhances context capture, which is crucial for medical image segmentation.

\begin{figure}
	\includegraphics[width=1.0\textwidth]{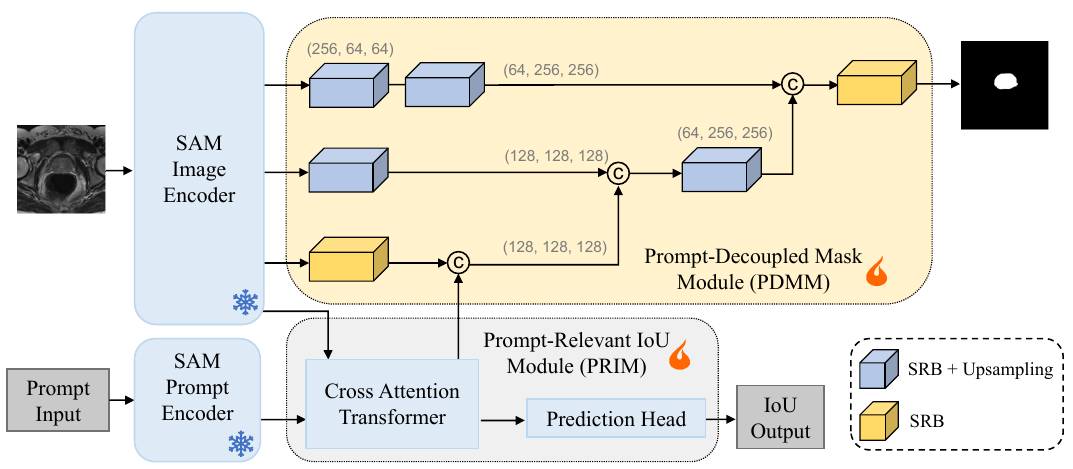}
	\centering
	\caption{Overview of the proposed DeSAM. The DeSAM consists of the image and prompt encoders of SAM, a prompt-decoupled mask module (PDMM), and a prompt-relevant IoU module (PRIM). The image encoder are used to compute the image  embeddings before training. The prompt encoder is frozen during training. The PRIM consists of a cross-attention transformer and an IoU prediction head, and it utilizes the image and prompt embeddings to generate mask embeddings and IoU score. The PDMM contains multiple channel attention-based residual blocks (SRB) and upsampling operations, and it integrates the mask embeddings and image embeddings to generate the mask.} 
	\label{fig1}
\end{figure}

\section{Decoupled Segment Anything Model}
\subsection{Architecture}
The overview of DeSAM is illustrated in Fig. \ref{fig1}. In addition to the encoder inherited from SAM, DeSAM contains two main components, the prompt-relevant IoU module (PRIM) and the prompt-decoupled mask module (PDMM). These components are described in detail in this section.

\subsubsection{Prompt-Relevant IoU Module (PRIM).}
The PRIM has a similar structure to the mask decoder of SAM, which includes a cross-attention transformer layer and an IoU prediction head. However, to decouple the prompt and the output mask, we only discard the mask prediction head and extract mask embeddings from the cross-attention transformer layer. 

\subsubsection{Prompt-Decoupled Mask Module (PDMM).}
PDMM is a crucial component of the proposed DeSAM architecture, designed to generate the final segmentation mask by fusing multi-scale image embeddings with prompt-relevant mask embeddings. Inspired by the success of U-Net \cite{ronneberger2015u} and UNETR \cite{hatamizadeh2022unetr} in medical image segmentation, we adopt a similar encoder-decoder structure for PDMM.

The first step in PDMM is to extract multi-scale image embeddings from the SAM ViT-H image encoder. Specifically, we select the output features from the global attention layers i = (8, 16, 24), which have a spatial resolution of 1280×64×64. These intermediate features capture hierarchical representations of the input image at different scales, providing rich contextual information for the subsequent mask-generation process.

To efficiently process and refine the extracted image embeddings, we employ a series of squeeze-and-excitation (SE) residual blocks \cite{hu2018squeeze}, followed by upsampling operations. The number of SE residual blocks applied to each image embedding varies depending on its spatial resolution. The refined image embeddings are then upsampled to match the resolution of the final segmentation mask. To further improve the information flow and gradient propagation within PDMM, we introduce skip connections that fuse the upsampled image embeddings from different scales. This fusion strategy allows the network to combine low-level spatial details with high-level semantic information, enabling more precise and detailed segmentation masks.

Finally, we merge the mask embeddings generated by the PRIM with the bottleneck embeddings of PDMM. This fusion serves two essential purposes. First, it allows PDMM to leverage the pre-trained weights of SAM's mask decoder, which encodes valuable knowledge about object shapes and boundaries. Second, it ensures a smooth gradient flow between PRIM and PDMM during training, facilitating effective end-to-end optimization of the entire DeSAM architecture.

By decoupling the mask generation process from the prompt embeddings and introducing a dedicated module for multi-scale image feature fusion, PDMM significantly enhances the ability of DeSAM to generate accurate and robust segmentation masks in a domain-agnostic manner.

\subsection{Training strategies}
During training, we load the pre-trained weights of SAM ViT-H, freeze the image and prompt encoders, and fine-tune the layers within PDMM and PRIM.
Since the automatic segmentation includes the grid points mode and the whole box mode, we adopt two different strategies to train the proposed model. In the grid points mode, we randomly generate points within and outside the ground truth mask in a 1:1 ratio. Mask generation is supervised by dice loss $L_{dice}$ and cross-entropy loss $L_{ce}$. IoU is supervised by mean square error loss $L_{mse}$. The total loss is:

\begin{equation}
	\label{eq1}
	\begin{small}
		L_{points} = \lambda_1 L_{dice} + \lambda_2 L_{ce} + \lambda_3 L_{mse}
	\end{small}
\end{equation}

The loss weight $\lambda_1$,  $\lambda_2$,  $\lambda_3$ are 1, 1, and 10. In the whole box mode, since the ground truth mask must be inside the box, we only supervise the mask generation:

\begin{equation}
	\label{eq1}
	\begin{small}
		L_{box} = L_{dice} + L_{ce}
	\end{small}
\end{equation}

\begin{figure}
	\includegraphics[width=1.0\textwidth]{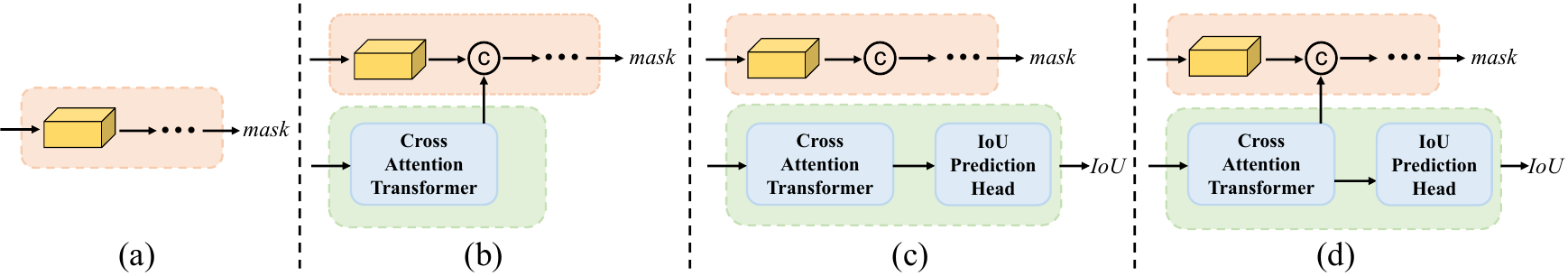}
	\centering
	\caption{Design choices of the decoder. (a) Generating a mask by directly using the image embedding from the encoder (PDMM only). (b) PDMM and PRIM without IoU prediction head. (c) PDMM and PRIM without mask embedding fusion. (d) Our proposed DeSAM.} 
	\label{fig3}
\end{figure}

\section{Results and Discussion}

\subsection{Dataset and implementation details}
To evaluate the performance of the proposed DeSAM, we conducted assessments in two cross-domain settings: 1) cross-modality abdominal multi-organ segmentation and 2) cross-site prostate segmentation. For cross-modality abdominal segmentation, we used two publicly available datasets \cite{landman2015miccai,kavur2021chaos}. For the multi-site prostate segmentation dataset, we collected three publicly available datasets from six different clinical sites, including NCI-ISBI-2013 \cite{NCI-ISBI_2013}, I2CVB \cite{lemaitre2015i2cvb}, and PROMISE12 \cite{litjens2014promise}. We adopted the same preprocessing method as MaxStyle \cite{chen2022maxstyle}. In each experiment, we split the data from one domain into a training dataset and an in-domain validation set in a 9:1 ratio. Then, we tested the robustness of the other domains using the best model on the in-domain validation set. We use the dice score as the evaluation metric to measure the quality of the predicted masks.

\begin{table}[t]
	\caption{Quantitative results of our ablation experiments. The best performance is indicated by bolded fonts. IPH: IoU prediction head. MEF: mask embedding fusion.}
	\label{tab1}
	\centering
	\resizebox{1.\textwidth}{!}{%
		\begin{tabular}{ccccc|ccccccc}
			\toprule
			\multicolumn{5}{c|}{Methods}                                                                              & \multicolumn{7}{c}{Dice (\%)}                                                                                          \\ \midrule
			\multicolumn{1}{c|}{Settings}     & PDMM       & PRIM       & IPH & MEF & A              & B              & C              & D              & E              & F              & Overall        \\ \midrule
			\multicolumn{1}{c|}{1}            & \checkmark &            &                     &                       & 79.09          & 74.96          & 54.19          & 80.11          & 77.22          & 77.53          & 73.85          \\
			\multicolumn{1}{c|}{2}            & \checkmark & \checkmark &                     & \checkmark            & 76.74          & 76.61          & 58.94          & 80.45          & 78.39          & 79.56          & 75.12          \\
			\multicolumn{1}{c|}{3}            & \checkmark & \checkmark & \checkmark          &                       & 79.52          & 78.37          & 59.59          & 82.65          & 80.16          & 74.58          & 75.81          \\
			\multicolumn{1}{c|}{4 (proposed)} & \checkmark & \checkmark & \checkmark          & \checkmark            & \textbf{82.80} & \textbf{80.61} & \textbf{64.77} & \textbf{83.41} & \textbf{80.36} & \textbf{82.17} & \textbf{79.02} \\ \bottomrule
		\end{tabular}%
	}
\end{table}

The image embeddings were precomputed before training using the ViT-H model as image encoder. We set the number of points for the grid points mode to 9x9. For network optimization, we used a learning rate of 1e-4 with a batch size of 8 and applied learning rate decay. The network was trained for 50 epochs for the prostate dataset to ensure convergence. We performed the experiments on a single RTX 3060 12GB. During training, the video memory usage was approximately 7.8 GB. We conducted the system-level comparison of our method with the upper bound (fully supervised on the seen domain), baseline (no domain generalization), MedSAM \cite{ma2024segment}, SAMed \cite{zhang2023customized} and other state-of-the-art single-source domain generalization methods, including 1)  adversarial bias field (AdvBias) \cite{chen2020advbias}; 2) RandConv \cite{xu2021randconv}; 3) MaxStyle \cite{chen2022maxstyle}; and 4) causality-inspired domain generalization (CSDG) \cite{ouyang2022csdg}. Upper bound and baseline experiments were implemented using nnU-Net \cite{isensee2021nnu}, and all other comparison methods used their recommended settings. We use two variants of DeSAM: DeSAM-B, which uses a whole image bounding box as the prompt, and DeSAM-P, which employs a grid of points as the prompt.

\begin{figure}
	\includegraphics[width=1.0\textwidth]{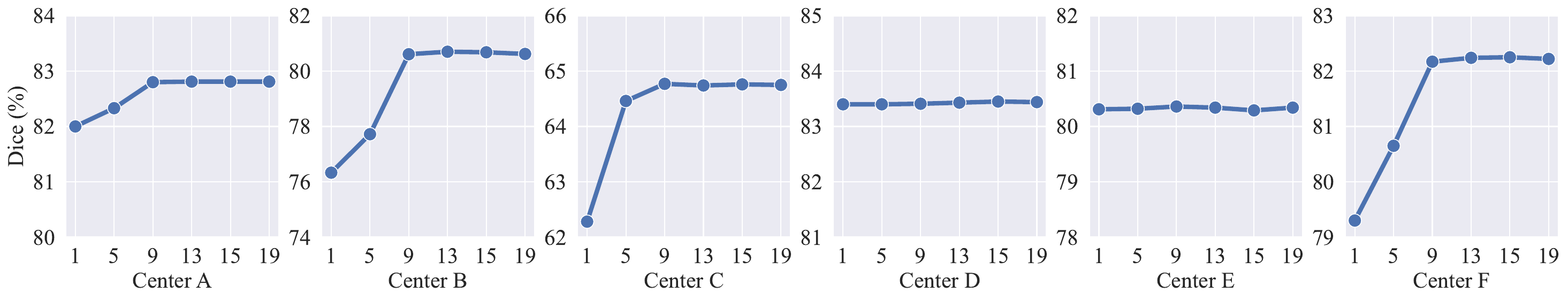}
	\centering
	\caption{Quantitative results of different number of points.} 
	\label{fig4}
\end{figure}

\begin{table}[t]
	\caption{Quantitative comparison of our DeSAM and state-of-the-art single-source domain generalization methods. The best performance is indicated by bolded fonts, and the second-best results is underlined.}
	\label{tab2}
	\centering
	\resizebox{1.\textwidth}{!}{%
		\begin{tabular}{c|cc|ccccccc}
			\toprule
			\textbf{}                         & \multicolumn{2}{c|}{Abdominal}    & \multicolumn{7}{c}{Prostate}                                                                                          \\
			Method                            & CT              & MRI             & A              & B              & C              & D              & E              & F              & Overall         \\ \midrule
			Upper bound \cite{isensee2021nnu} & 91.89           & 88.78           & 85.38          & 83.68          & 82.15          & 85.21          & 87.04          & 84.29          & 84.63           \\
			Baseline \cite{isensee2021nnu}    & 70.13           & 66.36           & 63.73          & 61.21          & 27.41          & 34.36          & 44.10          & 61.70          & 48.75           \\ \midrule
			AdvBias\cite{chen2020advbias}     & 75.04           & 74.20           & 77.45          & 62.12          & 51.09          & 70.20          & 51.12          & 50.69          & 60.45           \\
			RandConv\cite{xu2021randconv}     & 78.92           & 73.41           & 75.52          & 57.23          & 44.21          & 61.27          & 49.98          & 54.21          & 57.07           \\
			MaxStyle\cite{chen2022maxstyle}   & 82.92           & 76.93           & 81.25          & 70.27          & 62.09          & 58.18          & 70.04          & 67.77          & 68.27           \\
			CSDG\cite{ouyang2022csdg}         & 83.57           & 77.54           & 80.72          & 68.00          & 59.78          & 72.40          & 68.67          & 70.78          & 70.06           \\ \midrule
			MedSAM\cite{ma2024segment}        & 80.64           & 72.10           & 72.32          & 73.31          & 61.53          & 64.46          & 68.89          & 61.39          & 66.98           \\
			SAMed \cite{zhang2023customized}  & 77.21           & 70.35           & 73.61          & 75.89          & 58.61          & 73.91          & 66.52          & 72.85          & 70.23           \\
			DeSAM-B                           & {\ul 84.87}     & {\ul 79.57}     & {\ul 82.30}    & {\ul 78.06}    & \textbf{66.65} & {\ul 82.87}    & {\ul 77.58}    & {\ul 79.05}    & {\ul 77.75}     \\
			DeSAM-P                           & \textbf{86.68} & \textbf{80.05} & \textbf{82.80} & \textbf{80.61} & {\ul 64.77}    & \textbf{83.41} & \textbf{80.36} & \textbf{82.17} & \textbf{79.02} \\ \bottomrule
		\end{tabular}%
	}
\end{table}

\subsection{Ablation studies}
We performed a series of ablation experiments to validate the key design choices in our DeSAM on the prostate segmentation dataset. If not specified, the grid points mode with 9×9 points are used for prediction by default setting. For the ablation study, we compared our DeSAM with three variants: (a) Generating a mask by directly using the image embedding from the encoder (PDMM only). (b) PDMM and PRIM without IoU prediction head. (c) PDMM and PRIM without mask embedding fusion. Fig. \ref{fig3} shows the difference between the three variants and DeSAM. The results in Table \ref{tab1} show that the performance of our model is improved by gradually adding these components.

To investigate the robustness of our method in the grid points mode, we experimented with different numbers of grid points. Fig. \ref{fig4} reports the quantitative results. The results show that there is no degradation in the performance of the model as the number of grid points increases, indicating the effectiveness of our method in avoiding false positive masks affected by poor prompts in the grid points mode.

\begin{figure}
	\includegraphics[width=1.0\textwidth]{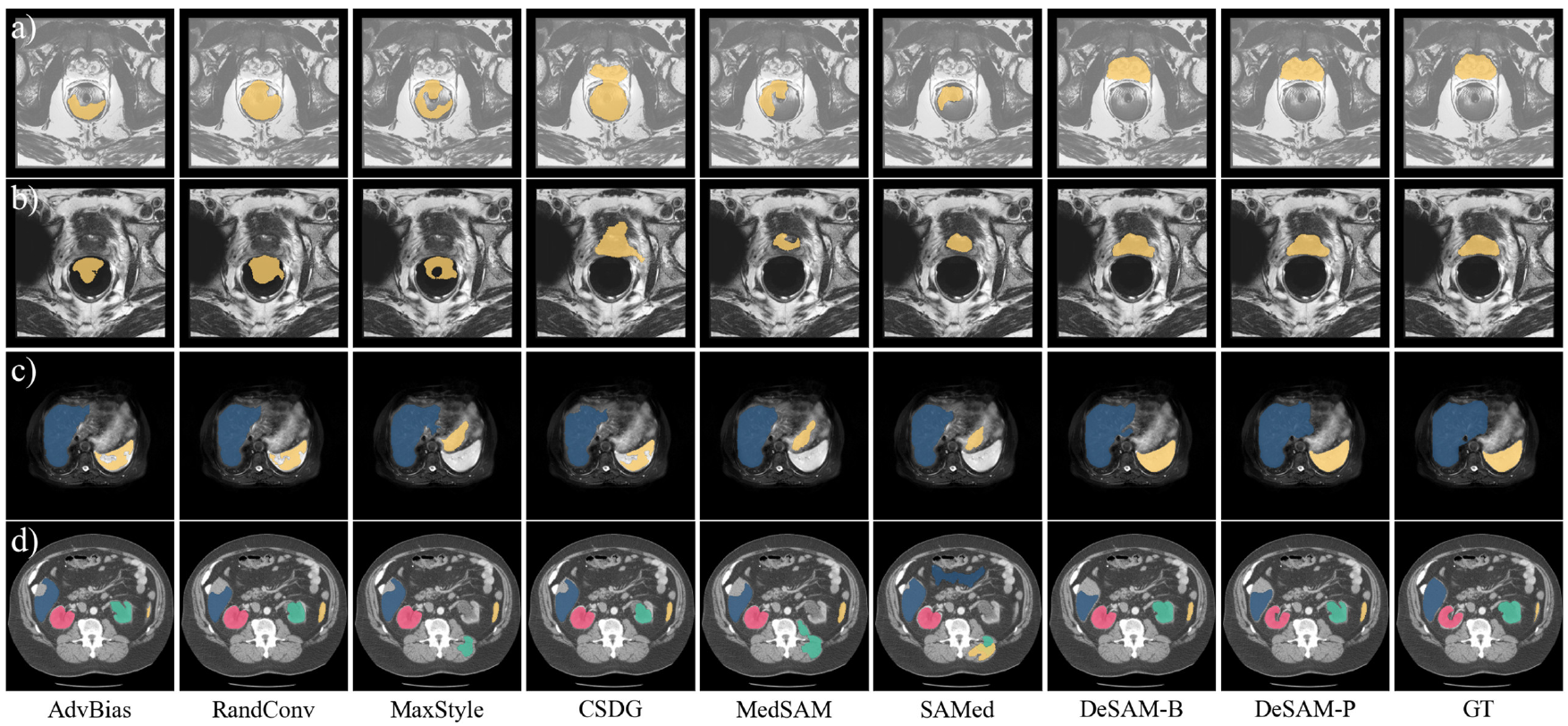}
	\centering
	\caption{Visual comparison of different methods for cross-site prostate segmentation and cross-modality abdominal multi-organ segmentation. GT represents the ground truth.} 
	\label{fig2}
\end{figure}

\subsection{Comparison with state-of-the-art methods}
The experimental results on both abdominal and prostate datasets are presented in Table \ref{tab2}. DeSAM-P achieves the best overall performance with a Dice score of 79.02\%, outperforming all other methods by a significant margin. Notably, DeSAM-P surpasses the previous state-of-the-art method, CSDG, by 8.96\% on the overall Dice score. These results highlight the effectiveness of our decoupling strategy in improving the generalization ability of SAM-based models for medical image segmentation. On the abdominal dataset, DeSAM-P and DeSAM-B consistently outperform other methods on both CT and MRI modalities. DeSAM-P achieves Dice scores of 86.68\% and 80.05\% on CT and MRI, respectively, setting new state-of-the-art performance levels. Fig. \ref{fig2} shows a visual comparison between the different methods of prostate segmentation. In contrast, our DeSAM achieves better results than the others, with no false positives in the background and segmentation boundaries very close to the ground truth.

\section{Conclusion}
We introduce DeSAM, a powerful network architecture for single-source domain generalization in medical image segmention. It decouples mask generation from prompt and takes advantage of the pre-trained weights of SAM. DeSAM motivates the decoder to learn prompt-invariant features from robust image embeddings. Moreover, DeSAM has strong ability to resist unseen distribution changes by fusing image embeddings at multiple scales. We validated the performance of DeSAM on two public datasets, demonstrating that the proposed method outperforms other state-of-the-art methods.

\begin{credits}
	\subsubsection{\ackname} This work was supported in part by National Science Foundation of China under Grant 82372052, in part by Taishan Industrial Experts Program under Grant tscx202312131, in part by Key Research and Development Program of Shandong under Grant 2021SFGC0104, in part by Science Foundation of Shandong under Grant ZR2022QF071 and ZR2022QF099, and in part by Key Research and Development Program of Jiangsu under Grant BE2021663 and BE2023714.
	
	\subsubsection{\discintname}
	The authors have no competing interests to declare that are relevant to the content of this article.
\end{credits}

\bibliography{mybibliography}
	
\end{document}